\documentclass[twocolumn,secnumarabic,amssymb, nobibnotes, aps, prl]{revtex4-1}

\usepackage{booktabs}

\usepackage{amsmath}
\usepackage{slashed}
\usepackage{mathtools}
\usepackage{mathrsfs}

\usepackage[utf8]{inputenc}
\usepackage{graphicx}
\graphicspath{  {images/} }

\begin{document}

\title{The curious case of the double-slit experiment and a black hole}%

\author{Satish Ramakrishna}
\email{ramakrishna@physics.rutgers.edu}
\affiliation{Department of Physics, Rutgers The State University of New Jersey, 136 Frelinghuysen Road, New Brunswick NJ 08854}

\author{Onuttom Narayan}
\email{onarayan@ucsc.edu}
\affiliation{Department of Physics, University of California, Santa Cruz, CA 95064}

\date{\today}
\maketitle
\tableofcontents

\section{Introduction}
This experiment was conceived of as a method of transmitting information from inside a black hole to the outside. As it turns out, it doesn't work in the form described (and possibly not in any form), but the way in which Nature prevents quantum-mechanical effects from transmitting usable information using quantum correlations is illuminating. In the process, one can learn some quantum theory, as well as quantum optics.

The proposed scheme uses a double-slit experiment, in the manner of the Delayed Choice set up \cite{Scully1} \cite{Scully2}, where the region where the interference takes place (between ``signal'' photons)  is spatially separated from the region where the Delayed Choice (with ``idler'' photons) is made. Indeed, this Double-Delayed Choice, which is this thought experiment, has one of the idler photons slip inside the event horizon and serves as the method to attempt to communicate from the inside to the outside. 

To briefly summarize the experiment: the thought experiment involves setting up a Double-Delayed-Choice set-up, with the ``signal'' and one ``idler'' photon outside and one ``idler'' photon inside the black hole, with the interference being observed outside the black hole. The information about which path the photons take before being detected  (henceforth referred to as the ``which-path information'') is independently determined by the observer inside as well as outside the black hole. The observer inside the black hole (whom we shall refer to as ``Babu'') has the opportunity to determine the which-path information for the photons independently of the observer outside (whom we shall refer to as ``Alisha'').  If Babu's actions can influence Alisha's observations, then Babu (sitting inside the black hole's event horizon)  can effectively communicate with Alisha outside the black hole.

\section{The Delayed Choice Experiment}
\subsection{The classic double-slit experiment}

The classic double-slit experiment \cite{Feynman3} is carried out by shooting electrons at a pair of closely placed slits - the electron flux is sufficiently small that one is able to count the number of times electrons hit various points on a screen placed past the slits. If no measures are taken to identify which of the two paths the electrons actually took to reach the screen, then the probability density of arrival at various points on the screen displays an ``interference'' pattern. If, however, the experiment is set up so as to identify which slit the electron went through, for example by shining an intense beam of photons at the slits that scatter off the electrons, then the interference pattern for those ``which-path'' identified electrons switches to a ``clump'' pattern, centered between the two slits. The standard experiment is displayed, schematically, in Fig. 1.

\begin{figure}[h]
\caption{The classic double-slit experiment, performed in free space}
\centering
\includegraphics[scale=1]{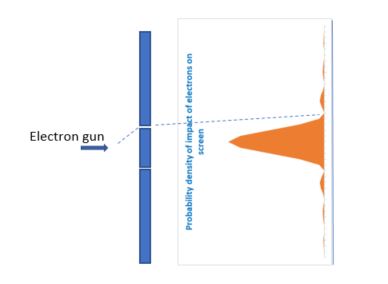}
\end{figure}

As an aside, an interesting variant of this experiment can be conceived where the slits are respectively inside and outside the event horizon of a classical black hole. In that case, following the usual principles of quantum mechanics and assuming that entanglement between the two paths for the electron survives transit through the event horizon for one of the paths,  the observer outside the black hole measures a "clump" pattern, while the observer inside measures an "interference" pattern. This is because while electrons can fall into the event horizon of a classical black-hole, they cannot escape from inside it; hence the outside observer can only get electrons from one of the two slits, i.e., only from the one outside the event horizon. The observer inside the event horizon can of course receive electrons that pass through both slits. This experiment is displayed, schematically, in Fig. 2.
\vspace{1 mm}

\begin{figure}[h]
\caption{The classic double-slit experiment, performed near a black hole}
\centering
\includegraphics[scale=1]{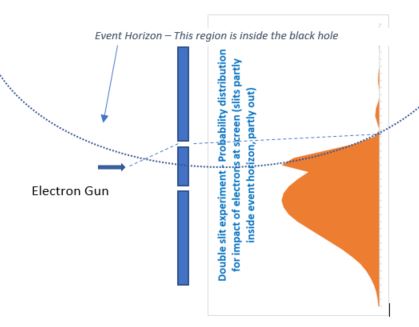}
\end{figure}

It is interesting to see what happens if we were to separate the isolation of path information from that of the interference phenomenon. That is done in the Delayed-Choice set up \cite{Wheeler} \cite{Scully1} \cite{Scully2} and is described in the next section (see Fig. 3)

\subsection{Double-slit with Delayed-Choice}

\begin{figure}[h]
\caption{The Delayed-Choice experiment}
\centering
\includegraphics[scale=.75]{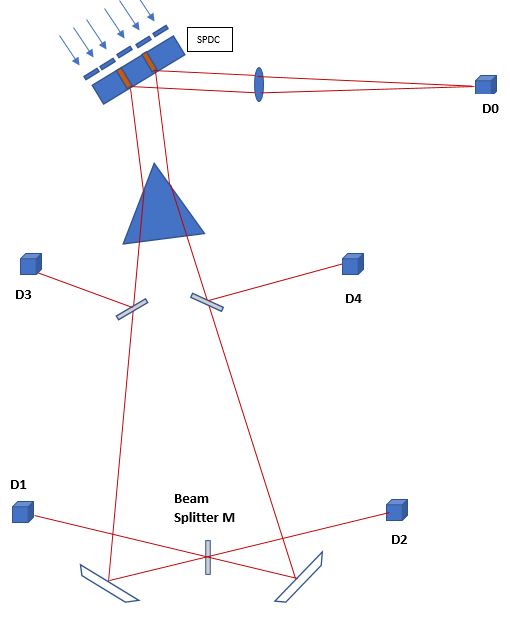}
\end{figure}

The delayed-choice experiment was proposed by Scully et al \cite{Scully1} and is portrayed in the accompanying Fig. 3. An SPDC (spontaneous parametric down conversion) setup that produces two photons, when one is incident on the inside face of the non-linear optical crystal, is used (this is marked as $SPDC$ in the figure). The photons that travel towards the interferometer/detector $D0$ are referred to as ``signal'' photons. The photons that travel towards the prism are the ``idler'' photons. After passing through the prism, the ``idler'' photons pass through a beam-splitter, that has a $50 \%$ chance of deflecting the incoming photon to detectors $D3$ and $D4$ respectively and $50\%$ probability of letting the photon pass on to the fully silvered (reflecting mirrors) at the bottom of the picture. Another beam-splitter is placed between the detectors $D1$ and $D2$,  so photons that are detected at $D1$ and $D2$ have their ``which-path'' information obliterated - for instance, an ``idler'' photon arriving at $D1$ could have come along either of two paths. The actual experiment was performed by Kim et al \cite{Scully2}.

 The detector $D0$ accumulates ``signal'' photons - a coincidence counter correlates it to ``idler'' photons detected at the detectors $D3, D4, D1$ and $D2$ (the ``idler'' photons arrive at those detectors a few nanoseconds after the ``signal'' photons are received. From the accumulated ``signal'' photons received at $D0$, if we separate the ones received in coincidence with the detectors $D3$ or $D4$, since the ``which-path'' information is clear in those cases, the pattern of interference observed (in the spatial density of the ``signal'' photons) is the ``clump'' pattern. However, the accumulated ``signal'' photons received at $D0$ that are coincident with the ones received at $D1$ display an interference pattern, since one cannot infer the path taken by the ``idler'' photons that show up at detector $D1$, which means one cannot tell which slit the ``signal'' photon came through before arriving at detector $D0$. Similarly, the accumulated ``signal'' photons received at $D0$ that are coincident with the ones received at $D2$ display an interference pattern in their spatial distribution, however this pattern is spatially offset $\pi$ (half a wavelength) off the one due to $D1$ - this is due to the unitarity principle that applies at the beam-splitter \cite{Scully1}. Briefly, the outgoing states from a beam-splitter are related to the incoming states by a unitary S-matrix, which, for a 50\% reflectance beam-splitter should be 

\begin{equation}
\left( \begin{array}{cc}
\frac {1}{\sqrt{2}} \	\frac {1}{\sqrt{2}} \\
-\frac {1}{\sqrt{2}}  \hspace{3mm}		\frac {1}{\sqrt{2}}   \end{array} \right)
\end{equation}
The relative minus sign, which is required to make the determinant unity (conservation of probability) is what leads to the phase-shifted interference pattern and is intrinsic to path-erasure experiments (see the analysis in the last section).

More generally, the transformation can be 
\begin{equation}
\left( \begin{array}{cc}
\alpha\	\hspace{3mm}	\beta\\
-\beta^*  \hspace{3mm} \alpha   \end{array} \right)
\end{equation}

 Notice, crucially, that the observer of the ``idler'' photons could choose to remove the beam-splitter (M) - this makes the path information immediately clear to this observer. The result is that the interference of the ``signal'' photons is eliminated (since the path of the photons has been identified); tracing this back, this is because one knows which slit that individual ``signal'' photon has traveled through on its way to $D0$.

\section{The Double-Delayed Choice setup}

The main contribution of this note is the Double-Delayed-Choice setup. It is described in Fig. 4.

\begin{figure}[h]
\caption{The Double-Delayed-Choice experiment}
\centering
\includegraphics[scale=.45]{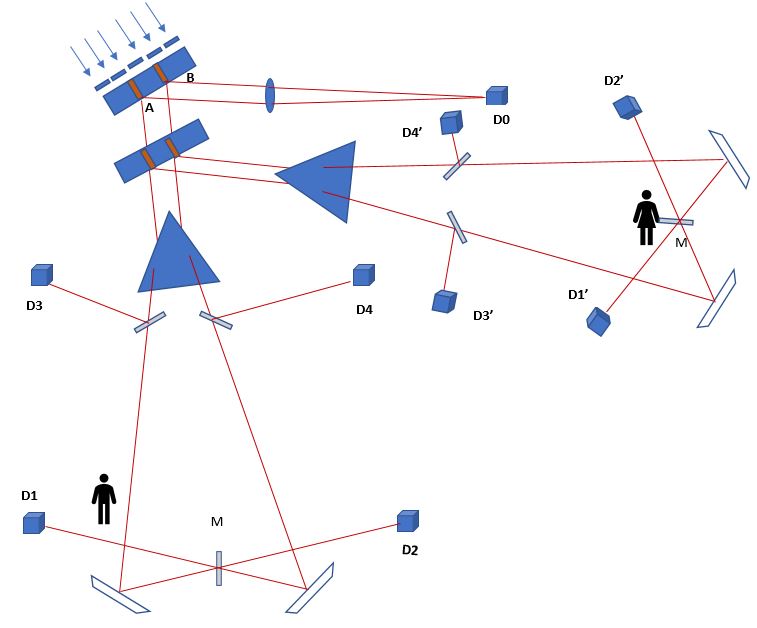}
\end{figure}

The crucial modification in this setup is that the ``idler'' photons pass through yet another SPDC (or similar non-linear) crystal set-up, to produce two ``child-idler'' photons. One of the ``child-idler'' photons is sent down to Babu and Alisha, who are observers of each leg of the ``child-idlers''. Note that if a ``child-idler'' is detected at one of the detectors $D1$ or $D2$ by Babu, at the same time, Alisha will detect a ``child-idler'' at $D1'$ or $D2'$, since two ``child-idler'' photons are produced simultaneously at the frequency doubler. Yet again, ``child-idlers'' that get detected by $D3, D4, D3', D4'$ clearly reveal their ``which-path'' information, so the corresponding ``signal' photons exhibit a ``clump'' pattern at the interferometer/detector $D0$ - notice that if even one of the detectors $D3, D4, D3', D4$ are triggered, immediately that particular ``signal'' photon will contribute to a ``clump'' pattern at $D0$. If neither of the ``child-idler'' photons trigger  $D3, D4, D3', D4'$, then clearly only the detectors $D1, D2, D1', D2'$ are triggered.  If either Babu and Alisha detect photons at $D1, D2, D1', D2'$, then if Alisha (for instance) analyzes the spatial density of ``signal'' photons corresponding to (and coincident with)  ``child-idler'' photons detected at $D1'$, we might expect that she would discover an interference pattern in the ``signal'' photons at $D0$.

A possible refutation of the general picture is the no-cloning theorem, which states that there cannot exist a machine that makes copies of arbitrary quantum states. We offer two reasons why this does not apply here. First, note, that the states of the ``idler'' photons are not ``arbitrary'' - the idlers come from one slit or the other and the state is specified as a linear superposition of the two states. The no-cloning theorem does not preclude copying precisely specified states again and again - for example, this is why stimulated emission does not violate the no-cloning theorem. Next, the input states (the ``idler'' came from either slit) are orthogonal, so an equal-weighted linear combination of these states can indeed be cloned. The limitations due to monogamy of entanglement also do not apply, since in terms of the paths (coming out of one of the two slits), the two ``child-idler'' photons are in the \underline{same} state - indeed the state of the entire system can be written as a $GHZ$ state,
$\frac{|111>+|222>}{\sqrt{2}}$ 
where  $|1>$ and $|2>$ refer to choice of the first or the second slit to pass through for the ``signal'' as well as the two ``child-idler'' photons. What's been done here is to set up a $GHZ$ state, separate the third particle from the first two, interfere the first two and make observations on the third particle.

So here is the potential scheme - If Babu decides to remove the beam-splitter M from his leg of the experiment, then the ``which-path'' information for his ``child-idler'' photon, hence of the ``signal'' photon,  is exposed. Now, the interference pattern that Alisha (we think) detected with the ``signal'' photons corresponding to her $D1'$ detection will vanish, to be replaced by a ``clump'' pattern. Thus by analyzing the interference pattern with corresponding ``signal'' photons, Alisha can deduce that Babu has removed his beam-splitter M. By inserting and removing his beam-splitter for a successive sequence of a large number of  photons, Babu can communicate, in a fashion, with Alisha.

Continuiing, suppose Babu were to alternate between inserting (1) and removing (0) his half-silvered mirror for every sequential bunch of N  ``child-idler'' photons. The pattern that Alisha observes with the ``signal'' photons corresponding to detections at $D1'$ and $D2'$ will either be of the ``interference'' type or ``clump'' type. The question reduces to how many observations of the spatial distribution does one need to establish that one is seeing an ``interference'' pattern or a ``clump'' pattern. Using standard principles from signal-processing, one needs to sample the distribution at double the maximum spatial frequency - so for all practical purposes, one needs $N \sim 2 L/a$ samples, where $L$ is the width of the screen and $a$ is the slit separation. The standard error in the patterns will be of $\it{\Theta}({\sqrt{N}})$, which corresponds to a relative error of $\frac{1}{\it{\Theta}({\sqrt{N}})}$. As $N \rightarrow \infty$, the accuracy of the message transmission gets closer to certainty.

There are, of course,  practical difficulties with this set-up. The SPDC conversion process is very inefficient, so a rather large number of initial photons needs to be produced $(\sim 10^{12})$ in order to produce one ``signal'' and ``idler'' pair. A factor of the same amount would need to be multiplied for the second SPDC setup to produce one pair of ``child-idler'' photons. And to communicate reasonably, we'd need a large number $(N)$ of ``child-idler''s.

\section{The Double-Delayed-Choice setup near a black hole}

\begin{figure}[h]
\caption{The Double-Delayed-Choice experiment near a black hole}
\centering
\includegraphics[scale=.45]{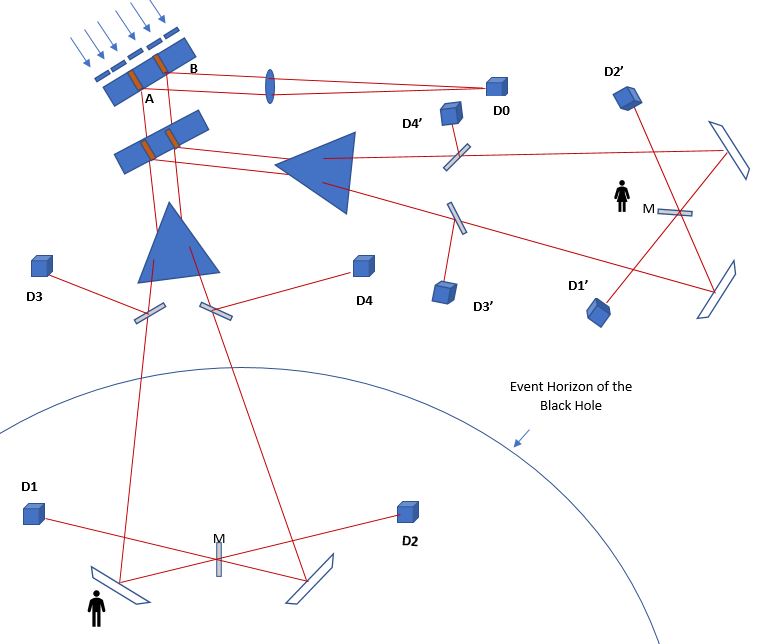}
\end{figure}

The next step is to arrange things so that Babu is inside the event horizon of a black hole. Again, if this scheme were to work, Babu could arrange to insert or remove the beam-splitter M for sequences of photons, to communicate imperfectly with Alisha. 

\section{This is wrong!}

So, it turns out that this is incorrect. Alisha will not, after all, see an interference pattern for the signal photons coincident with $D1'$ or $D2'$. In fact, as we shall see, an interference pattern is only seen for signal photons coincident with one of Alisha's detectors as well as one of Babu's detectors. If we ignore the results of one of the observers, the pattern we will see for the interference of the signal photons will be purely of the "clump" sort. To obtain this, we need to follow Feynman's famous paraphrasing of the quantum prescription - "you add up the amplitudes and it gives you the wave-function". 

\section{Formal analysis}

Let's start by considering the regular Delayed-Choice experiment. For simplicity, in all that follows, we can ignore the detectors $D_3, D_4, D_{3'}, D_{4'}$ since they do not contribute to interference, though they are an unavoidable by-product of the experimental set-up that was used by Kim et. al. \cite{Scully2}.

Let's write down the states for the system as follows -  $|S_A>, |S_B>$ represent the states of the signal photon (there are two slits, $A$ and $B$ that it could emerge out of) and $|I_A>, |I_B>$ represent the two idler photons. $|D_1>, |D_2>$ represent the states of the idler photon post the beam splitter (corresponding to each of the two detectors). In general, we must have the states entering and departing the beam-splitter related through a general unitary transformation (for $2 X 2$ matrices).
\begin{eqnarray*}
\begin{aligned}
\left( \begin{array}{c}
|I_A>\\
|I_B>  \end{array} \right) = 
\left( \begin{array}{cc}
\alpha\	\hspace{3mm}	\beta\\
-\beta^*  \hspace{3mm} \alpha   \end{array} \right) 
\left( \begin{array}{c}
|D_1>\\
|D_2>  \end{array} \right) \\
\end{aligned}
\end{eqnarray*}

Post the SPDC set-up, the states of the photons can be represented as 
$\frac {|S_A>|I_A>+|S_B>|I_B>}{\sqrt{2}}$.
Continuing, post the beam-splitter, the state is
\begin{equation}
\frac {|S_A>(\alpha |D_1> + \beta |D_2> )+|S_B>(-\beta^{*} |D_1> + \alpha^{*} |D_2> )}{\sqrt{2}}
\end{equation}
which we can re-organize into
\begin{equation}
\frac {|D_1>(\alpha |S_A> - \beta^{*} |S_B> )+|D_2>(\beta |S_A> + \alpha^{*} |S_B> )}{\sqrt{2}}
\end{equation}

If we consider coincident hits with $D_1$ and the signal photons, we are asking for the matrix element with $<D_1|$ and that is
$\frac{(\alpha |S_A> - \beta^{*} |S_B> )}{\sqrt{2}}$ . If we now asked for the position representation of this state, along the x-direction (perpendicular to the beam of signal photons), we'd get an interference term in the intensity proportional to $- (\alpha \beta + \alpha^{*} \beta^{*})$. In addition, the interference term from the other detector (derived from looking at the matrix element with $<D_2|$, that would be proportional to $(\alpha \beta + \alpha^{*} \beta^{*})$. Note that if we simply looked for the interference term without considering which detector (out of $D_1$ and $D_2$) was triggered, we'd need to add the above interference terms - the coefficients are equal and opposite in sign, so they cancel. We'd only see interference if we specifically broke down the events to those signal photon events coincident with detector $D_1$ alone and then, separately, those signal photon events coincident with detector $D_2$ alone.

Next, let's consider the Double-Delayed Choice set-up. Let's write down the states for the system as follows -  $|S_A>, |S_B>$ represent the states of the signal photon (there are two slits, $A$ and $B$ that it could emerge out of) and $|I^{1}_A>, |I^{1'}_A>, |I^{1}_B>, |I^{1'}_B>$ represent the staes accessible to the two idler photons. $|D_1>, |D_{1'},>|D_2>, |D_{2'}>$ represent the states of the respective idler photons post the beam splitter (on each of the two arms $1$ and $1'$, corresponding to each of the two detectors). 
Here, we write the initial state (post the two SPDC set-ups) as $\frac {|S_A>|I_1>|I_{1'}>+|S_B>|I_2>|I_{2'}>}{\sqrt{2}}$. 

In general, the states entering and departing the beam-splitter must be related by a general unitary transformation (for $2 X 2$ matrices).
\begin{eqnarray*}
\begin{aligned}
\left( \begin{array}{c}
|I^{1}_A>\\
|I^{1}_B>  \end{array} \right) = 
\left( \begin{array}{cc}
\alpha\	\hspace{3mm}	\beta\\
-\beta^*  \hspace{3mm} \alpha   \end{array} \right) 
\left( \begin{array}{c}
|D_1>\\
|D_2>  \end{array} \right) \\
\left( \begin{array}{c}
|I^{1'}_A>\\
|I^{1'}_B>  \end{array} \right) = 
\left( \begin{array}{cc}
\alpha'\	\hspace{3mm}	\beta'\\
-\beta^{'*}  \hspace{3mm} \alpha'   \end{array} \right) 
\left( \begin{array}{c}
|D_{1'}>\\
|D_{2'}>  \end{array} \right)
\end{aligned}
\end{eqnarray*}

the state after the beam splitters is

\begin{equation}
\begin{aligned}
\frac {1}{\sqrt{2}} (|S_A>(\alpha |D_1> + \beta |D_2> )(\alpha' |D_{1'}> + \beta' |D_{2'}> ) \\
+|S_B>(-\beta^{*} |D_1> + \alpha^{*} |D_2> )(- \beta'^* |D_{1'}> + \alpha'^* |D_{2'}> ) )
\end{aligned}
\end{equation}
which we can re-organize into
\begin{equation}
\begin{aligned}
\frac {1}{\sqrt{2}} |D_{1'}> [ \alpha' |S_A> (\alpha |D_1> + \beta |D_2>) \\
- \beta'^* |S_B> (- \beta^* |D_1> + \alpha^* |D_2> )  ] \\
+\frac {1}{\sqrt{2}}  |D_{2'}>[ \beta' |S_A> (\alpha |D_1> + \beta |D_2>) \\
+ \alpha'^* |S_B> (- \beta^* |D_1> + \alpha^* |D_2> )  ]
\end{aligned}
\end{equation}

Again, let's take the matrix element vs $<D_{1'}|$; this is the state that we need to work with if we consider detections coincident with $D_{1'}$ (Alisha's detector).
\begin{equation}
\begin{aligned}
\frac {1}{\sqrt{2}} (\alpha \alpha' |S_A> + \beta^* \beta'^* |S_B>) |D_1>  \\
+ \frac {1}{\sqrt{2}} (\alpha' \beta |S_A> - \alpha^* \beta'^* |S_B>) |D_2>
\end{aligned}
\end{equation}
and similarly, the matrix element with $<D_{2'}|$
\begin{equation}
\begin{aligned}
\frac {1}{\sqrt{2}} (\alpha \beta' |S_A> - \alpha'^* \beta^* |S_B>) |D_1>  \\
+ \frac {1}{\sqrt{2}} (\beta \beta' |S_A> +\alpha^* \alpha'^* |S_B>) |D_2>
\end{aligned}
\end{equation}

It is then easy to see that the interference term for the triggering of \\
$D_{1'}$ and $D_1$ is proportional to $(\alpha \alpha' \beta \beta' + c.c.)$ \\
$D_{1'}$ and $D_2$ is proportional to $- (\alpha \beta' \beta \alpha' + c.c.) $ \\
$D_{2'}$ and $D_1$ is proportional to $- (\beta' \alpha \alpha' \beta + c.c.) $ \\
$D_{2'}$ and $D_2$ is proportional to $(\beta \beta' \alpha \alpha' + c.c.)$ \\
where "c.c." represents the complex conjugate of the preceding term.
If, for instance, we look only at the interference of signal photons coincident with detector $D_{1'}$ without paying attention to the detectors $D_1$ and $D_2$, then we need to add the first two terms. This term is ZERO. So, no interference is seen, regardless of the magnitudes of $\alpha, \beta, \alpha', \beta'$. 

So, unless we look very precisely at the coincident signal and child-idler detections for all three detectors, we will not see an interference pattern. Sitting outside a black hole, Alisha will not be able to tell that there is ANY quantum interference without splitting those events into the ones coincident with her individual detectors as well as the ones inside the black hole (four combinations). "Tracing" over Babu's detectors will simply give her no interference. And since she cannot communicate with Babu inside the black hole, there is no chance that she can have access to the events that Babu sees - ergo, \underline{she sees no interference whatsoever}.

There is a traditional quantum optics way of carrying out this analysis, which we write along the lines of Kim et. al. \cite{Scully2}. For simplicity, we will perform this calculation for the simple unitary beam-splitter matrix with $\alpha = \beta = \frac {1}{2}$.
The joint detection counting rate $R_{0jk}$ of detector $D0$, $D_j$ (belonging to Babu) and detector $D_k$ (belonging to Alisha) is given by an extension of the Glauber formula \cite{Glauber} to
\begin{eqnarray*}
\begin{aligned}
R_{0jk} \propto \frac {1}{T} \int_0^T \int_0^T \int_0^T dT_0 dT_j dT_k  \\
<\Psi|E_0^{(-)}E_j^{(-)}E_k^{(-)}E_k^{(+)}E_j^{(+)}E_0^{(+)}|\Psi> \\
= \frac {1}{T} \int_0^T \int_0^T \int_0^T dT_0 dT_j dT_k   |<\Psi|E_k^{(+)}E_j^{(+)}E_0^{(+)}|\Psi>|^2
\end{aligned}
\end{eqnarray*}
where we define the state (as in the previous analysis) as
\begin{equation}
|\Psi>= \sum_{s, i_1, i_2} C(\vec{k}_s,\vec {k}_{i_1}, \vec {k}_{i_2})  a^{\dagger}_s(\omega(\vec{k}_s)a^{\dagger}_{i_1}(\omega(\vec {k}_{i_1}) a^{\dagger}_{i_2}(\omega(\vec {k}_{i_2})
\end{equation}
and $C(\vec{k}_s,\vec {k}_{i_1}, \vec {k}_{i_2})  = \delta(\omega(\vec{k}_s)+\omega(\vec {k}_{i_1})+\omega(\vec {k}_{i_2})-\omega_{pump})\delta(\vec{k}_s,\vec {k}_{i_1}, \vec {k}_{i_2}- \vec{k}_{pump})$ for the doubled SPDC. The frequencies and wave-vectors are, labeled for the "signal", "child-idler"1 (Babu's photon) and "child-idler"2 (Alisha's photon). We consider again, the amplitudes for joint detection

\begin{eqnarray*}
\begin{aligned}
\psi(t_0, t_j^{i_1}, t_k^{i_2})=<0|E_k^{(+)}E_j^{(+)}E_0^{(+)}|\Psi>
\end{aligned}
\end{eqnarray*}
where $t_0=T_0 - L_0/c, t_j^{i_{1,2}}=T_j^{i_{1,2}}-L_j^{i_{1,2}}/c$ for each of the detectors in the paths, $j=1,1',2,2',3,3',4,4'$ and the $L$'s are the optical distances from the output points of the initial BBO crystal to the respective detectors.These optical distances are used to compute travel times along the paths followed by the photons. There are now four possible amplitudes for the coincident detection of photons where the "which-path" information is erased, i.e., in detectors $1$ and $2$, due to the slits $A$ and $B$, along the lines of \cite{Scully2}. Note that in the below notation, for example, $t_1^{1A}$ refers to detection of "child-idler"-1 photon at detector $1$ ("Babu"'s detector) when it emerged from slit $A$. Also, $\mathscr{A}(t_0, t_i^{1A})$ is, for instance, the amplitude $<0|E_i^{1}|A>$ for Babu's "child-idler" photon. These ampliudes are computed by the simple product of the amplitudes (taken with sign after the beam-splitter) for the two orthogonal paths the photons can take to the beam-splitter.
\begin{eqnarray*}
\begin{aligned}
\psi(t_0, t_1^{1}, t_{1'}^{2})=(\mathscr{A}(t_0, t_1^{1A})  \mathscr{A}(t_0, t_{1'}^{2A}) \\
+ \mathscr{A}(t_0, t_1^{1B}) \mathscr{A}(t_0, t_{1'}^{2B})) \\
\psi(t_0, t_1^{1}, t_{2'}^{2})=(\mathscr{A}(t_0, t_1^{1A}) \mathscr{A}(t_0, t_{2'}^{2A}) \\
- \mathscr{A}(t_0, t_1^{1B})\mathscr{A}(t_0, t_{2'}^{2B}) ) \\
\psi(t_0, t_2^{1}, t_{1'}^{2})=(- \mathscr{A}(t_0, t_2^{1A}) \mathscr{A}(t_0, t_{1'}^{2A}) \\
+ \mathscr{A}(t_0, t_2^{1B}) \mathscr{A}(t_0, t_{1'}^{2B})) \\
\psi(t_0, t_2^{1}, t_{2'}^{2})=(\mathscr{A}(t_0, t_2^{1A}) \mathscr{A}(t_0, t_{2'}^{2A}) \\
- \mathscr{A}(t_0, t_2^{1B})  \mathscr{A}(t_0, t_{2'}^{2B}))
\end{aligned}
\end{eqnarray*}
In addition we have, for the detectors $3$ and $4$ where the "which-path" information is known (if detector $3$ was triggered, the photon clearly came from slit $A$!)
\begin{eqnarray*}
\begin{aligned}
\psi(t_0, t_1^{1}, t_{3'}^{2})=\mathscr{A}(t_0, t_1^{1A})  \mathscr{A}(t_0, t_{3'}^{2A})  \\
\psi(t_0, t_1^{1}, t_{4'}^{2})= \mathscr{A}(t_0, t_1^{1B})  \mathscr{A}(t_0, t_{4'}^{2B})  \\
\psi(t_0, t_3^{1}, t_{1'}^{2})=\mathscr{A}(t_0, t_3^{1A})  \mathscr{A}(t_0, t_{1'}^{2A})  \\
\psi(t_0, t_4^{1}, t_{1'}^{2})=\mathscr{A}(t_0, t_4^{1B})  \mathscr{A}(t_0, t_{1'}^{2B})  \\
\end{aligned}
\end{eqnarray*}

Making the same approximations as in \cite{Scully2} for waveform shapes so that the integrals are simplified,
\begin{equation}
 \mathscr{A}(t_0, t_i^{Q1}) = \mathscr{A}_0  \Pi(t_0-t_i^{Q1}) e^{- i \Omega_0 t_0} e^{- i \Omega_i t_i^{Q1}} 
\end{equation}

with 
\begin{equation}
\begin{aligned}
 \Pi(t_0-t_i^{1})  = 
\left(  1 \hspace{3 mm} :  \hspace{3 mm}0 < t_0-t_i^{1} < D_L \right)
\end{aligned}
\end{equation}
and $ \Pi(t_0-t_i^{1}) $ is zero if the condition is not met. Since we need to multiply the pulse functions due to each "child-idler" with the "signal" photon,  the three photons need to originate (for entanglement purposes) from a common piece of optical medium originally.
Putting this together, we obtain the coincidence counting rates  (note $x$ is the transverse dimension where the incidence of the "signal" photons is observed, $f$ is the focal distance from the BBO crystal's focusing lens to the screen where the "signal" photons are observed and $\lambda$ is the wavelength of the "signal" photons).
\begin{equation}
\begin{aligned}
R_{011} \propto cos^2(2\pi x d/\lambda f) \\
R_{012} \propto sin^2(2\pi x d/\lambda f) \\
R_{021} \propto sin^2(2\pi x d/\lambda f) \\
R_{022} \propto cos^2(2\pi x d/\lambda f) \\
\end{aligned}
\end{equation}

and the other coincidence counting rates for the other cases have no interference terms.

Note that if we are indifferent to the particular detector that was triggered ($1$ or $2$) in for Babu's "child-idler", then 
$R_{011}+R_{012} \propto 1$ which means that the "signal" photon displays no interference whether or not both "child-idler" photons have their "which-path" information erased. This is consistent with our earlier analysis. Unless we actually separate the signal photon events into those corresponding to individual detectors belonging to Babu \underline{and} Alisha, we will not see any interference type pattern in the signal photon's spatial distribution.

\section{Conclusions}

This experiment is designed to exploit the GHZ state to elicit correlations that can be used to transmit information. The delicate cancellations due to quantum mechanics are a subtle reminder of how these long range correlations predicted by the quantum theory prevent actual information from being transmitted beyond the light cone.

\section{Acknowledgments}

SR acknowledges useful discussions of various aspects with Scott Thomas, Vijay Jain, Pouya Asadi and S. Govindarajan.


\end{document}